\documentclass[prd,onecolumn,preprintnumbers,floats,floatfix,apsrev,amsmath,amsfonts,nofootinbib,superscriptaddress]{revtex4-2}
\usepackage{color}
\usepackage{graphicx}
\usepackage{dcolumn}
\usepackage{natbib}
\usepackage{hyperref}
\usepackage{hypernat}
\usepackage{natbib}
\usepackage{soul}
\usepackage{amssymb}

\newcommand{\dd}{{\rm d}}

\newcommand{\p}{\partial}

\newcommand{\be}{\begin{equation}}
\newcommand{\en}{\end{equation}}

\newcommand{\paren}[1]{\left({#1}\right)}

\begin{document}
\title{Scattering and absorption of massless scalar waves by a ModMax black hole}

\author{Arturo Quispe Baptista}
\email{aquispeb@fcpn.edu.bo}
\affiliation{Carrera de F\'{i}sica, Facultad de Ciencias Puras y Naturales, UMSA, La Paz, Bolivia.}

\author{M.~L.~Pe\~{n}afiel}
\email{miguelpenafiel@upb.edu}
\affiliation{UERJ - Universidade do Estado do Rio de Janeiro 150, CEP 20550-013, Rio de Janeiro, RJ, Brazil.}
\affiliation{FIA - Facultad de Ingeniería y Arquitectura, Universidad Privada Boliviana, Camino Achocalla Km 3.5, La Paz, Bolivia.}


\begin{abstract}
ModMax electrodynamics is a remarkable example of nonlinear electrodynamics that preserves both conformal and duality invariance. When coupled to General Relativity, the resulting black hole solutions introduce a tunable nonlinearity parameter that effectively screens the electromagnetic charge. We study the scattering and absorption of massless scalar waves by a ModMax black hole by applying the partial waves method. We also compute some well-known analytical approximations for the scattering and absorption cross sections and contrast them with the full numerical solution. Our results adequately reproduce those of Maxwell electrodynamics in the appropriate limit and illustrate the effect of the charge screening in ModMax electrodynamics.
\end{abstract}

\maketitle

\section{Introduction}

Black holes (BHs) are among the most interesting predictions of general relativity. Despite the theory being highly nonlinear, BHs represent very simple solutions that can be characterized by up to three parameters: namely their mass, angular momentum per unit mass, and charge. The study of how BHs can interact with different matter fields is of particular interest, being the absorption and scattering properties of different kinds of waves of particular relevance due to the possible observational consequences. In this context, the absorption and scattering properties of scalar fields by a Schwarzschild BH have been studied long ago (see, for instance, \cite{Matzner1968,Collins1973,Sanchez1976,Sanchez1978a,Andersson1995}). Several generalizations including other types of BH solutions and matter fields have been studied as well \cite{Mashhoon1973,Fabbri1975,Glampedakis2001,Dolan2006,Crispino2007,Crispino2009,Dolan2008,Crispino2009a,Crispino2009b,Benone2014}, providing insight into how such  phenomena can help us distinguish between different BH configurations. Notwithstanding, it is interesting to note that some universal results hold for the absorption cross section of different types of BHs \cite{Decanini2011}.

A possible direct generalization for the study of wave scattering by BHs is to consider solutions coming from alternative theories of gravity \cite{Li2025}. Yet, it is also possible to consider the scattering by BHs that are solutions of general relativity coupled to alternative matter fields, such as nonlinear electrodynamics (NLEDs). Theories of NLED are often introduced as generalizations of Maxwell's electromagnetism in order to address certain situations that Maxwell's theory fails to recover, such as vacuum birrefringence or the possibility of obtaining regular field solutions \cite{Plebaǹski1970,Sorokin2021}. These theories have been extensively studied, Born-Infeld electrodynamics being the most representative example \cite{Born1934}. Another interesting example of NLED is the recently derived ModMax electrodynamics \cite{Bandos2020,Kosyakov2020}, which stands out as a theory that is both conformal and duality invariant and thus represents an interesting alternative for obtaining physically meaningful solutions.\footnote{A  partial list of interesting developments within ModMax theory includes Refs. \cite{Kruglov2021,Banerjee2022,Lechner2022,Neves2023,Sorge2024,Kuzenko2024,AyonBeato2024}.}

Remarkably, the coupling of NLEDs to general relativity produces BH solutions that can have a wide spectrum of interesting physical properties. For instance, it is possible to construct regular static, spherically symmetric BHs when considering NLED and general relativity \cite{AyonBeato1998,Bronnikov2001,Dymnikova2004} as well as the newly discovered possibility of obtaining rotating BHs within NLED \cite{GarciaDiaz2021,Breton2022,AyonBeato2024a,GalindoUriarte2024}. The solution of a BH corresponding to general relativity coupled to ModMax electrodynamics was presented in \cite{FloresAlfonso2021}, and several important features have been studied, including: the shadow cast by the ModMax BH \cite{Pantig2022}, the study of accelerated BHs within ModMax theory \cite{Barrientos2022}, the propagation of light in ModMax spacetime \cite{GuzmanHerrera2024,GuzmanHerrera2024a}, the ModMax generalization of Majumdar-Papapetrou spacetime \cite{Bokulic2025}, among others. 

The scattering and absorption of massless scalar waves by BHs have been studied within a plethora of solutions coming from the minimal coupling between GR and NLED (see, for instance, \cite{Macedo2014,Macedo2015,Sanchez2018,Paula2020,Paula2022,Paula2023}). In this work we focus on studying the absorption and scattering of massless scalar waves by the ModMax BH, which is a benchmark for further studies of propagation of distinct fields within this solution.

This paper is organized as follows: in Sec. \ref{sec:ModMaxED} we review the main features of ModMax NLED including the ModMax BH solution and the study of the null geodesics in this spacetime, we compute the classical scattering cross section via the geodesic analysis, as well as the glory approximation. In Sec. \ref{sec:PartialWaves} we introduce the partial wave method to obtain the scattering and absorption cross sections for massless scalar waves impinging upon a ModMax BH for different non-linearity regimes. Finally, in Sec. \ref{sec:concl} we give our final remarks and present some perspectives for future works. In the course of this work we use geometrized units and the metric signature is such that the Minkowski metric has the form $\eta_{\mu\nu}=\text{diag}\paren{+1,-1,-1,-1}$.

\section{ModMax electrodynamics} \label{sec:ModMaxED}

Maxwell's theory of electrodynamics is formulated as a $U(1)$ gauge theory, where the gauge field $A_\mu$  is the four-potential, and the antisymmetric Faraday tensor is given by $F_{\mu\nu}=\p_\mu A_\nu-\p_\nu A_\mu$ while its dual $\widetilde{F}^{\mu\nu}=\frac{1}{2}\eta^{\mu\nu\alpha\beta}F_{\alpha\beta}$, where $\eta^{\mu\nu\alpha\beta}$ is the totally antisymmetric Levi-Civita tensor.  Given such fields, it is possible to construct the two linearly independent invariants
\be \label{eq:invariants}
F\equiv\frac{1}{2}F^{\mu\nu}F_{\mu\nu}\quad   \text{and}\quad G\equiv\frac{1}{2}\widetilde{F}^{\mu\nu}F_{\mu\nu}\ .
\en
Accordingly, Maxwell's theory is defined by the Lagrangian density $\mathcal{L}_\text{M}=-F/8\pi$ and thus represents the only possible linear theory of electrodynamics. It is possible to generalize Maxwell's theory by considering a generic Lagrangian density which is a function of both invariants $F$ and $G$ in Eq. \eqref{eq:invariants} such that $\mathcal{L}=\mathcal{L}\paren{F,G}$, which we shall denote as nonlinear electrodynamics (NLED). The basic physical requirements that one may demand for a NLED are related to the recovery of Maxwell's theory at a certain limit, as well as obeying certain energy conditions \cite{Plebaǹski1970,Garcia1984}. Notwithstanding, in general, certain symmetries, such as duality and conformal symmetries need not be preserved in NLED, leaving the theories that preserve such symmetries as special cases of NLED.

\begin{figure}[b]
\centering
\includegraphics[width=0.45\columnwidth]{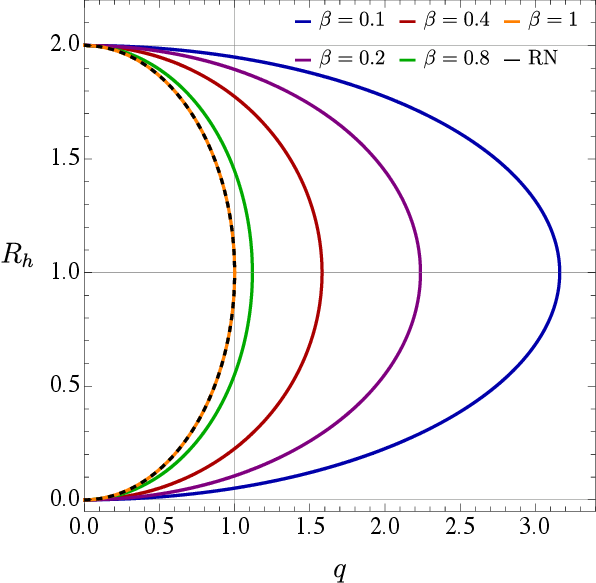}
\caption{Radius of the horizons, $R_h=r_h/M$ as a function of the reduced charge $q=\sqrt{Q_e^2+Q_m^2}/M$ for different values of $\beta=e^{-\gamma}$. The dashed line represents the horizons for the RN solution.}
\label{fig:horizons}
\end{figure}

One such theory that respects both duality and conformal invariance is ModMax electrodynamics \cite{Bandos2020,Kosyakov2020}, which is given by the Lagrangian density 
\be \label{eq:LModMax}
\mathcal{L}_{\text{ModMax}}=-\frac{F}{2}\cosh\gamma+\frac{1}{2}\sqrt{F^2+G^2}\sinh\gamma\ ,
\en
where $\gamma\ge0$ in order to respect causality and Maxwell electrodynamics is recovered in the limit $\gamma\to0$. This outstanding NLED respects both duality and conformal symmetries. In particular, conformal invariance requires the energy-momentum tensor of the theory being traceless, which can be stated by the requirement \cite{Kosyakov2007}
\be \label{eq:confinv}
F\mathcal{L}_F+G\mathcal{L}_G-\mathcal{L}=0\ ,
\en 
where $\mathcal{L}_X$ corresponds to the derivative of $\mathcal{L}$ with respect to a given invariant. On the other hand, duality invariance demands that
\be\label{eq:dualinv}
\widetilde{E}^{\mu\nu}E_{\mu\nu}=\widetilde{F}^{\mu\nu}F_{\mu\nu}\ ,
\en
where $E_{\mu\nu}=\p\mathcal{L}/\p F^{\mu\nu}=2\paren{\mathcal{L}_FF_{\mu\nu}+\mathcal{L}_G\widetilde{F}_{\mu\nu}}$ is the excitation tensor and $\widetilde{E}^{\mu\nu}$ its dual. Condition \eqref{eq:dualinv} can be translated to a differential equation involving $\mathcal{L}$ and its derivatives as
\be \label{eq:dualinv1}
G\left[4\paren{\mathcal{L}_F^2-\mathcal{L}_G^2}-1\right]=8F\mathcal{L}_F\mathcal{L}_G\ .
\en 
Direct inspection shows that the ModMax Lagrangian in Eq. \eqref{eq:LModMax} satisfies both conformal and duality invariance requirements in Eqs. \eqref{eq:confinv} and \eqref{eq:dualinv1}.

It is interesting to note that in the particular case of $G=0$, ModMax Lagrangian can be written as
\be
\mathcal{L}_{\text{ModMax}}=-\frac{F}{2}e^{-\gamma}=\mathcal{L}_{\text{Max}}e^{-\gamma}\ ,
\en
where $\mathcal{L}_{\text{Max}}$ is the Lagrangian density for Maxwell electrodynamics. This implies that, in the absence of magnetic fields, all Maxwell solutions can be directly mapped into ModMax electrodynamics \cite{Lechner2022}.
\subsection{ModMax black hole} \label{sec:ModMaxBH}
The minimal coupling of ModMax electrodynamics to Einstein's gravity leads to a static, spherically symmetric BH solution of mass $M$ and electric charge $Q_e$, whose line element is given by \cite{FloresAlfonso2021}
\be \label{eq:dsModMax}
\dd s^2=f\paren{r}\dd t^2-f\paren{r}^{-1}\dd r^2-r^2\dd\Omega^2\ ,
\en
where $\dd\Omega^2=\dd\theta^2+\sin^2\theta\dd\phi^2$ is the solid angle and
\be
f\paren{r}=1-\frac{2M}{r}+\frac{Q_e^2}{r^2}e^{-\gamma}\ .
\en
Due to the fact that ModMax electrodynamics preserves duality invariance, it is possible to perform a $SO(2)$ duality rotation between the fields and obtain the dyonic solution \cite{Bokulic2025}
\be
f_{\text{d}}\paren{r}=1-\frac{2M}{r}+\frac{\paren{Q_e^2+Q_m^2}}{r^2}e^{-\gamma}\ ,
\en
where $Q_m$ is the corresponding magnetic charge. Just as in the Reissner-Nordstr\"{o}m (RN) case, this solution exhibits an inner and outer horizon coming from the condition $f(r_\pm)=0$, and an extremal horizon that corresponds to the maximum allowable charge-to-mass ratio. Notice that the charge contribution is modulated by $e^{-\gamma}\le1$, which acts as a screening mechanism for the charge, and allows to have a larger charge-to-mass ratio than in the RN case before reaching extremality. Figure \ref{fig:horizons} displays the horizons $R_h=r_h/M$ for different configurations of $\beta=e^{-\gamma}$ and the reduced dyonic charge $q=\sqrt{Q_e^2+Q_m^2}/M$. It can be seen that for large $\gamma$, the screening effect on the charge is maximized and the allowable charge-to-mass ratio considerably increases.
\subsection{Geodesic scattering}
\begin{figure*}[ht]
\centering
\includegraphics[width=0.9\textwidth]{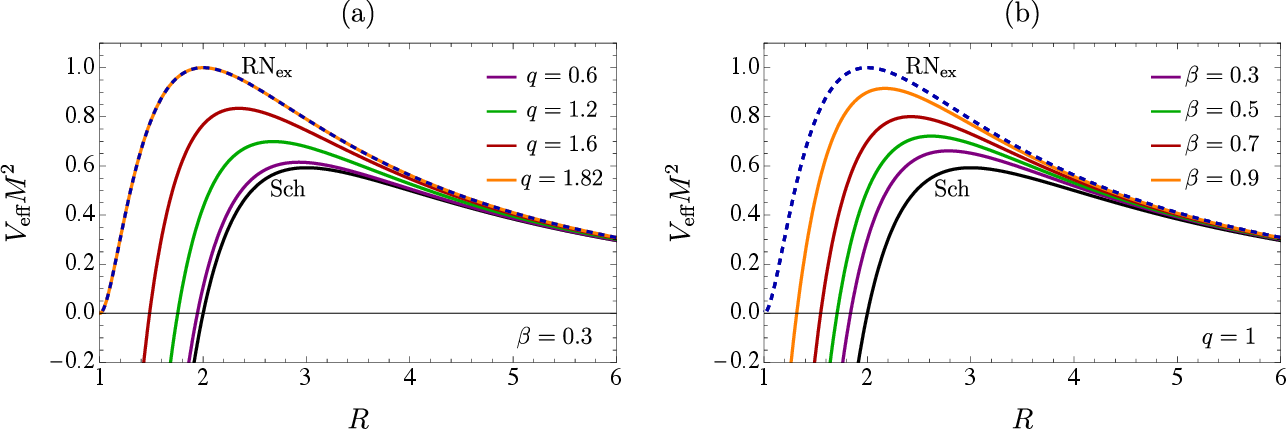}
\caption{Effective potential governing radial motion of massless particles for (a) fixed $\beta=e^{-\gamma}$ and variable $q$, and (b) fixed $q$ and variable $\beta$. The upper, blue dashed, curve corresponds to the effective potential of an extremal RN BH and the lower, black, curve corresponds to the effective potential of a Schwarzschild BH. All the possible effective potentials for the ModMax BH lie in between these curves.}
\label{fig:Veff}
\end{figure*}

We are interested in analyzing the null geodesics in this spacetime. Since the metric \eqref{eq:dsModMax} is static and spherically symmetric, there exist two Killing vectors, $\p_t$ and $\p_\phi$, that define the conserved energy and angular momentum of a test particle as
\be \label{eq:cons}
E=f\paren{r}\dot{t}\quad \ \text{and}\quad  L=r^2\dot{\phi}\ ,
\en 
where the overdot means derivative with respect to an affine parameter. Using Eq. \eqref{eq:cons} in \eqref{eq:dsModMax}, and setting $\theta=\pi/2$, we can write the equation of motion for massless particles in the form $\dot{r}^2+V_{\text{eff}}=E^2$, where 
\be \nonumber
V_{\text{eff}}=\frac{L^2}{r^2}f\paren{r}\ .
\en
Figure \ref{fig:Veff} displays the effective potential for massless particles in Schwarzchild, extremal RN and ModMax BHs for different charge-to-mass ratio and different values of $\beta=e^{-\gamma}$, respectively. As shown in the Figure, the extremal RN case possesses the highest maximum relative to the other cases. 

\begin{figure*}[ht]
\centering
\includegraphics[width=0.90\textwidth]{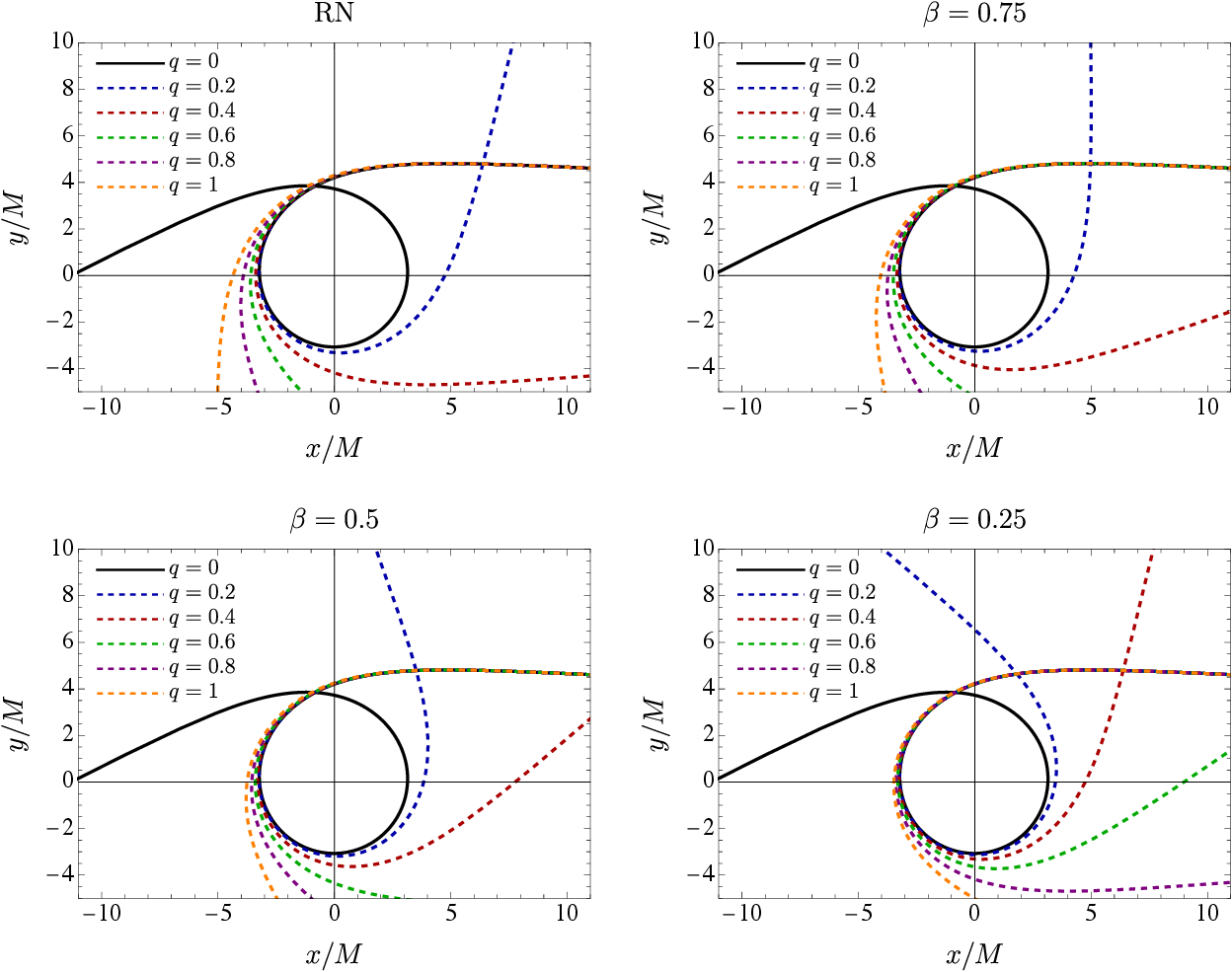}
\caption{Trajectories for massless particles (other than photons) in the ModMax spacetime for different values of $\beta=e^{-\gamma}$ and a fixed impact parameter of $b=5.2M$. The trajectories are found by numerically solving Eq. \eqref{eq:trajectories} with the adequate boundary conditions. We have set $r_{\infty}=100M$.}
\label{fig:gridplot}
\end{figure*}

We can rewrite the equation of motion by defining the impact parameter $b\equiv L/E$ and $u\equiv1/r$ as
\be\label{eq:eom1}
\paren{\frac{du}{d\phi}}^2=\frac{1}{b^2}-fu^2\ ,
\en
where $f=f(1/u)$. Taking the derivative with respect to $\phi$ of the latter equation leads to
\be\label{eq:trajectories}
\frac{d^2 u}{d\phi^2}=-\frac{u^2}{2}\frac{df}{du}-fu\ ,
\en
which can be solved by imposing the adequate boundary conditions. Figure \ref{fig:gridplot} displays the geodesics for massless particles approaching a ModMax BH for different configurations of $\beta=e^{-\gamma}$. It can be seen that, just as in the RN situation, as the BH's charge increases, the deflection angle diminishes. Also, as $\beta$ decreases, i.e., as the departure from Maxwell electrodynamics becomes steeper, the overall effect of the BH's charge is screened: massless particles impinging upon BHs with the same $q$ but different $\beta$ have different deflection angles.

\begin{figure*}[ht]
\centering
\includegraphics[width=0.9\textwidth]{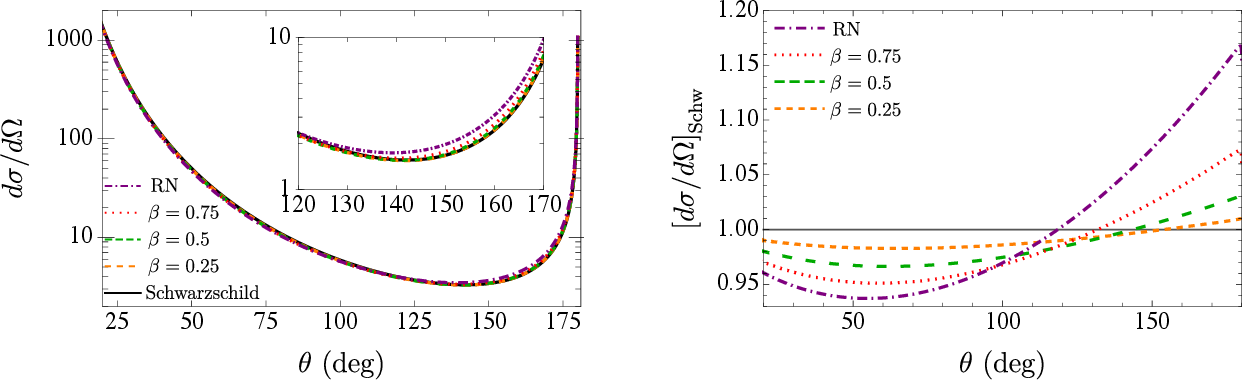}
\caption{(Left panel) Classical differential scattering cross section computed from Eq. \eqref{eq:classSigma} for $Q=1$ and different values of $\beta=e^{-\gamma}$, the Schwarzschild case is also displayed in the thick black curve. (Right panel) Classical differential scattering cross section normalized with respect to the Schwarzschild case, for different values of $\beta$ and $Q=1$. In all the above cases we have used $M=1$.}
\label{fig:rowscat}
\end{figure*}

The classical scattering cross section is given by
\be\label{eq:classSigma}
\left.\frac{d\sigma}{d\Omega}\right|_\text{class.}=\frac{1}{\sin\theta}b\paren{\varTheta}\left|\frac{db\paren{\varTheta}}{d\varTheta}\right|\ ,
\en
where $\theta$ is the scattering angle, and $\varTheta$ is the deflection angle. Both are related by $\varTheta=\theta+2n\pi$, with $n\in\mathbb{Z}^{+}$ being the number of times that the particle orbits the BH before being scattered to infinity. In order to evaluate the classical scattering cross section, we need to know the radius of closest approach of the geodesic, $r_0$. We begin by considering the case where we have an unstable null circular geodesic, for a critical radius $r_c=1/u_c$. This implies that the roots of Eqs. \eqref{eq:eom1} and \eqref{eq:trajectories} are zero, i.e., defining $f_c\equiv f(1/u_c)$, we have
\begin{align}
u_c b_c\sqrt{f_c}&=1\ , \label{eq:uc}\\
2f_c+u_c\frac{df_c}{du}&= 0\ ,
\end{align}
where $b_c$ is the impact parameter associated with the critical orbits. For $b>b_c$, the massless particles will be scattered by the BH, with a turning point $u_0$, that is the inverse of the maximum possible approximation of the particle, and is the largest value that $u$ can take for which Eq. \eqref{eq:eom1} vanishes. For such a case, we may integrate Eq. \eqref{eq:eom1} in order to obtain the deflection angle as
\be\label{eq:defang1}
\Theta\paren{b}=2\int_0^{u_0}\left[\frac{1}{b^2}-fu^2\right]^{-1/2}du-\pi\ .
\en
We can obtain the deflection angle in the weak field limit by expanding the integrand in Eq. \eqref{eq:defang1} for small $M$ and $Q$ and using Eq. \eqref{eq:uc} for the impact parameter. We can express the result in powers of $1/b$ as
\be
\Theta\paren{b}\approx\frac{4 M}{b}+\frac{3 \pi  \left(5 M^2-\beta  Q_e^2\right)}{4 b^2}+\mathcal{O}\paren{\frac{1}{b}}^3\ .
\en
Since $\beta\le1$, for a given $Q_e$, the deflection angle for the ModMax BH is smaller than the case of RN. This weak field approximation can be used to invert the relation between $\Theta$ and $b$ and directly compute the classical differential scattering cross section through Eq. \eqref{eq:classSigma}. Correspondingly, in order to find the classical scattering cross section in the general regime, the deflection angle must be obtained using \eqref{eq:defang1} and later inverted in order to express $b\paren{\Theta}$ such that the classical differential scattering cross section can be computed through Eq. \eqref{eq:classSigma}. Figure \ref{fig:rowscat} displays the classical differential scattering cross section for fixed values of $Q$ in different ModMax scenarios, namely $\beta=\paren{0.25,0.5,0.75,1}$. As shown in the Figure, ModMax electrodynamics effectively screens the charge, i.e., for a given charge, as ModMax departures from Maxwell's theory are more significant, the classical differential scattering cross section for a given BH becomes closer to the Schwarzschild case.  

\subsection{Glory scattering}
One may also consider the strong field regime where the deflection angle in Eq. \eqref{eq:defang1} is $\theta\sim\pi$, which corresponds to massless particles with energies close to the maximum of the effective potential \cite{Bozza2002}, giving rise to the glory effect. Such an effect anticipates interference phenomena that shall be present when treating the full partial waves solution for the problem. The semiclassical formula for the glory scattering by a spherically symmetric BH is given by \cite{Matzner1985}
\be \label{eq:glorySCS}
\left.\frac{d\sigma}{d\Omega}\right|_{\text{glory}}=2\pi\omega b_g^2\left|\frac{db}{d\theta}\right|_{\theta=\pi}\left[J_{2s}\paren{\omega b_g \sin\theta}\right]^2\ ,
\en
where $J_{2s}(x)$ is the Bessel function of the first kind and $s$ is the helicity of the scattered wave (hence $s=0$ corresponds to a scalar wave), $\omega$ is the frequency of the scalar wave and $b_g$ is the impact parameter of the backscattered wave. Notice that the deflection angle in Eq. \eqref{eq:defang1} takes into account the number of times that the particle orbits the BH before being (back)scattered, being the $n=0$ mode the responsible for the largest contribution to the glory SCS. 

\begin{figure}[b]
\centering
\includegraphics[width=0.45\columnwidth]{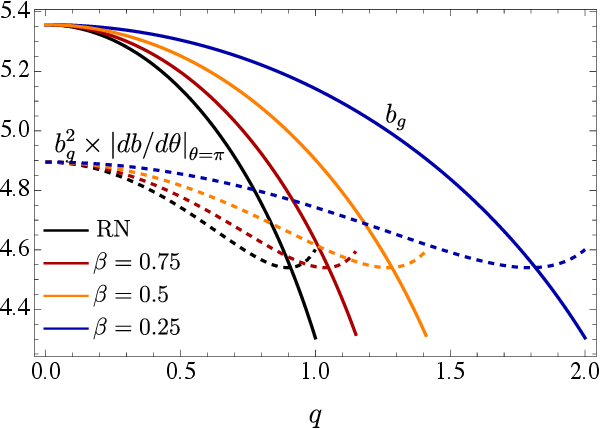}
\caption{Glory coefficients $b_g$ (thick lines) and $b_g^2\vert db/d\theta\vert_{\theta=\pi}$ (dashed lines) that enter into the glory SCS in Eq. \eqref{eq:glorySCS} for different values of $\beta=e^{-\gamma}$ as functions of $q$. It is evident that the coefficient $b_g$ is always a monotonically decreasing function of $q$, for any $\beta$, while $b^2_g\vert\ db/d\theta\vert_{\theta=\pi}$ possesses a local minimum.}
\label{fig:glorycoefs}
\end{figure}

Figure \ref{fig:glorycoefs} displays the glory parameter $b_g$ and $b_g^2\vert db/d\theta\vert_{\theta=\pi}$, which is related to the magnitude of the glory peak, for the $n=0$ mode in Eq. \eqref{eq:glorySCS} as a function of the normalized charge for different values of $\beta=e^{-\gamma}$. It can be seen that the parameter $b_g^2$ is always a monotonically decreasing function of $q$, irrespective of the value of $\beta$, while the glory peak possesses a local minimum for every value of $\beta$, just as in the Maxwell case (c.f. Ref. \cite{Crispino2009}). This local minimum for the glory peak has the same value for all ModMax scenarios, but it corresponds to a different charge configuration. The results for the glory peak are coincident with the screening mechanisms of the charge due to the electromagnetic nonlinearities.

\section{Partial waves method} \label{sec:PartialWaves}
We shall consider the propagation of a massless scalar field in the background of a ModMax BH, such that the scalar field obeys the Klein-Gordon equation
\be\label{eq:KG}
\nabla_\mu\nabla^\mu\Phi=\frac{1}{\sqrt{-g}}\p_\mu\paren{\sqrt{-g}g^{\mu\nu}\p_\nu\Phi}=0\ .
\en
Given the fact that the background spacetime is spherically symmetric, we can write
\be \label{eq:ansatz}
\Phi\paren{t,r,\theta,\varphi}=\frac{\psi_{\omega \ell}\paren{r}}{r}Y_{\ell m}\paren{\theta,\varphi}e^{-i\omega t}\ ,
\en
where $Y_{\ell m}\paren{\theta,\varphi}$ are the scalar spherical harmonics. Plugging this \emph{ansatz} into Eq. \eqref{eq:KG} results in the equation for the radial functions 
\be
f\paren{r}\frac{d}{dr}\paren{f\paren{r}\frac{d\psi_{\omega \ell}\paren{r}}{dr}}+\paren{\omega^2-V_{\text{eff}}\paren{r}}\psi_{\omega \ell}\paren{r}=0\ ,
\en
with $V_{\text{eff}}\paren{r}=f\paren{r}\left[\frac{l\paren{l+1}}{r^2}+\frac{f'\paren{r}}{r}\right]$. Furthermore, defining the tortoise coordinates as $f\paren{r}dr_{*}=dr$, we can rewrite the latter equation as 
\be \label{eq:Sch}
\frac{d^2}{dr_*^2}\psi_{\omega \ell}\paren{r_*}+\left[\omega^2-V_{\text{eff}}\paren{r_*}\right]\psi_{\omega \ell}\paren{r_*}=0\ .
\en
\begin{figure}[b]
\centering
\includegraphics[width=0.45\columnwidth]{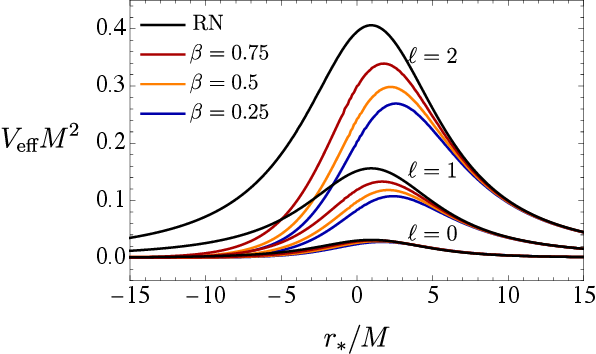}
\caption{Effective potential for $q=1$ and distinct values of $\ell$ for RN (black lines), $\beta=0.75$ (red lines), $\beta=0.5$ (orange lines) and $\beta=0.25$ (blue lines).}
\label{fig:pVeffbetas}
\end{figure}
Figure \ref{fig:pVeffbetas} displays the effective potential in tortoise coordinates for different values of $\ell$ and different ModMax scenarios for a fixed charge $q=Q/M$. It can be seen that the maximum for a given charge corresponds to the RN situation and it decreases as $\beta$ does, which is consistent with the charge being screened by the ModMax nonlinearities. Both near the horizon $(r_{*}\to-\infty)$ and at spatial infinity $(r_*\to+\infty)$ the effective potential goes to zero and the solution for Eq. \eqref{eq:Sch} in the $r\gg r_+$ regime is 
\be
\psi_{\omega \ell}\approx\omega r_*\left[(-i)^{\ell+1}A^\text{in}_{\omega\ell}h_\ell^{\paren{1}*}\paren{\omega r_*}+i^{\ell+1}A^{\text{out}}_{\omega\ell}h_{\ell}^{\paren{1}}\paren{\omega r_*}\right]\ ,
\en
where $h_\ell^{\paren{1}}\paren{x}$ are the spherical Hankel functions \cite{Abramowitz1964}, and $A^{\text{in}}_{\omega\ell}$ and $A^{\text{out}}_{\omega \ell}$ are complex constants. In the latter, the factor $\omega r_*$ in front ensures normalization for the solutions and, in the asymptotic regime we have $h_\ell^{\paren{1}}\paren{z}\approx(-i)^{\ell+1}e^{iz}/z$ for $z\to\infty$. Noticing that $V_{\text{eff}}\to0$ for $r_*\to\pm\infty$, we have
\be \label{eq:bcSch}
\psi_{\omega \ell}\paren{r_*}\approx\begin{cases}
A_{\omega \ell}^{\text{tr}}e^{-i\omega r_*}\ ,\ \paren{r_*\to-\infty}\ ,\\
A_{\omega\ell}^{\text{in}}e^{-i\omega r_*}+A_{\omega\ell}^{\text{out}}e^{i\omega r_*}\ ,\ \paren{r_*\to+\infty}\ ,
\end{cases}
\en
where the relation $\vert A_{\omega\ell}^{\text{in}}\vert^2=\vert A_{\omega\ell}^{\text{tr}}\vert^2+\vert A_{\omega\ell}^{\text{out}}\vert^2$ holds due to flux conservation.

\subsection{Scattering cross section}
\begin{figure*}[ht]
\centering
\includegraphics[width=0.9\textwidth]{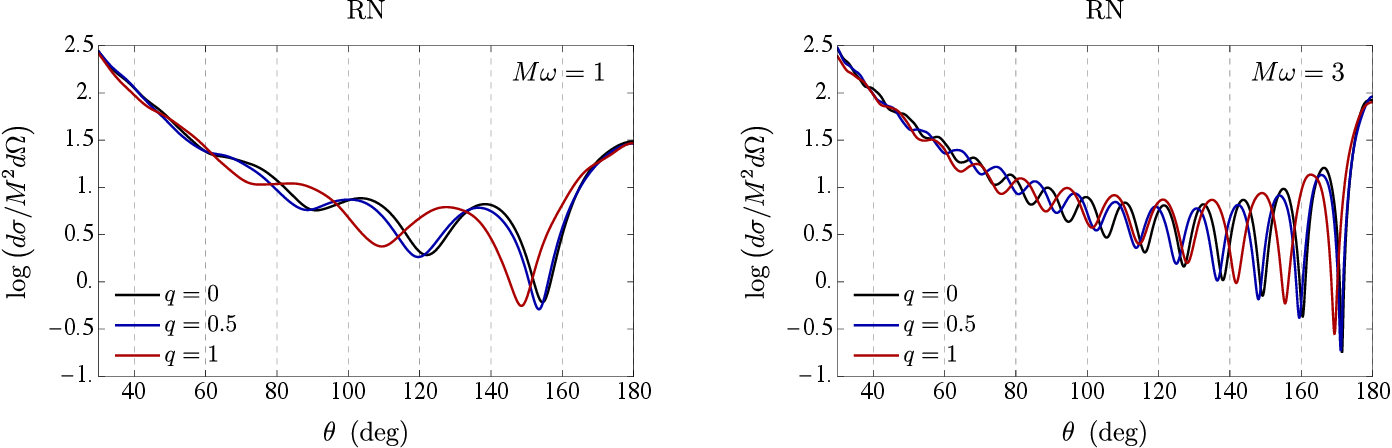}
\caption{Differential scattering cross section for RN spacetime for different values of $M\omega$ and $q=Q/M$.}
\label{fig:SCSrow}
\end{figure*}

The differential scattering cross section (SCS) is given by \cite{Sanchez1978}
\be \label{eq:SCSscalar}
\frac{d\sigma}{d\Omega}=\vert f\paren{\theta}\vert^2\ ,
\en
where 
\be \label{eq:fPl}
f\paren{\theta}=\frac{1}{2i\omega}\sum_{\ell=0}^{\infty}\paren{2\ell+1}\left[e^{2i\delta_\ell\paren{\omega}}-1\right]P_\ell\paren{\cos\theta}
\en
is the scattering amplitude and the phase shifts are
\be
e^{2i\delta_\ell\paren{\omega}}=\paren{-1}^{\ell+1}\frac{A^{\text{out}}_{\omega\ell}}{A^{\text{in}}_{\omega\ell}}\ .
\en
In order to calculate the differential SCS from Eq. \eqref{eq:SCSscalar}, it is necessary to solve the radial equation \eqref{eq:Sch} from a region very close to the horizon up to a region far away from the BH; in particular, we have chosen $r=r_h+10^{-5}$ and $r=1000r_h$, respectively. The coefficients $A_{\omega\ell}^{\text{in}}$ and $A_{\omega\ell}^{\text{out}}$ are obtained by matching the solution with the asymptotic behavior  given by Eq. \eqref{eq:bcSch}. The series in Eq. \eqref{eq:fPl} is poorly convergent due to the appearance of Legendre's polynomials. In order to circumvent this, we have used a regularizing procedure employed by Yennie \emph{et al.} \cite{Yennie1954} in the context of particle physics and by Dolan \emph{et al.} \cite{Dolan2006} in BH physics. This method consists on applying an iterative procedure to reduce the divergent behavior of the series by calculating new coefficients for the series, thus reducing the meaningful terms that must be taken into account when calculating the series in Eq. \eqref{eq:fPl}. We have tested our integration scheme by re-obtaining the well-known solutions present in the literature for the RN spacetime (c.f. \cite{Crispino2009,Sanchez2018}) (see Figure \ref{fig:SCSrow}).

\begin{figure*}[t]
\centering
\includegraphics[width=0.9\textwidth]{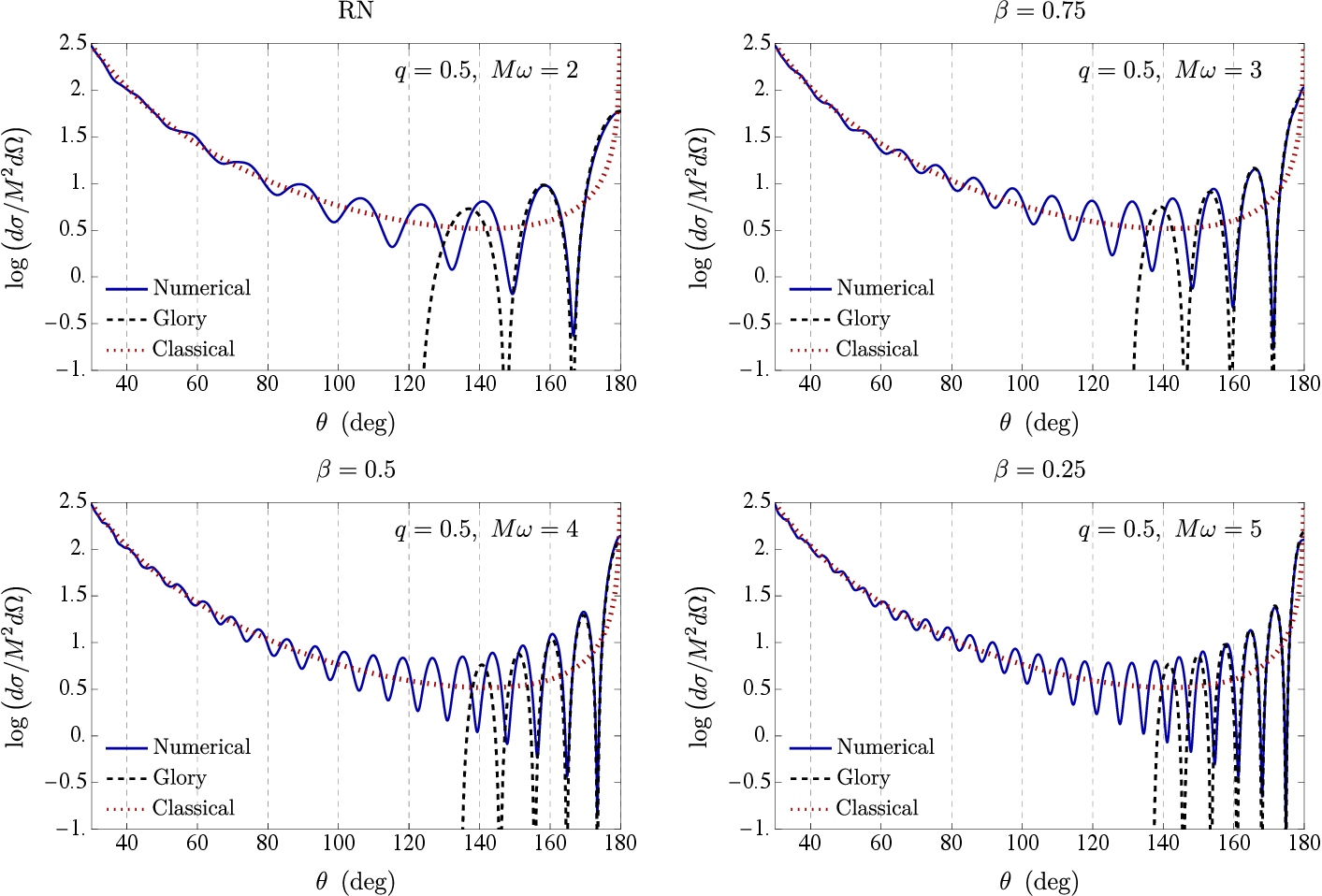}
\caption{Differential scattering cross section for RN spacetime for different values of $M\omega$ and $q=Q/M$.}
\label{fig:SCSgridcompact}
\end{figure*}

Figure \ref{fig:SCSgridcompact} displays the differential SCS obtained through the geodesic analysis, the glory approximation and the full partial waves solution for different scenarios of departures from Maxwell's theory. It can be seen that the classical approximation works well for small scattering angles and that the glory approximation effectively captures the behavior at large scattering angles, irrespective of the ModMax scenario. Let us stress that the effect of the charge screening due to ModMax nonlinearities is reflected in the fact that BHs corresponding to different configurations of $\beta$ and $q$ can display the same scattering features. 


\subsection{Absorption cross section} \label{sec:absCS}
\begin{figure}[b]
\centering
\includegraphics[width=0.45\columnwidth]{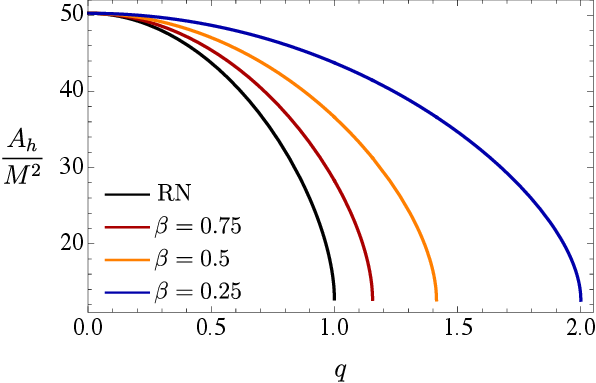}
\caption{Horizon area with respect to the normalized charge $q$ for distinct values of $\beta=e^{-\gamma}$. The effect of charge screening is evident: for a fixed charge, the area grows with diminishing $\beta$.}
\label{fig:phors}
\end{figure}

The total absorption cross section is given by
\be
\sigma_{\text{abs}}=\frac{\pi}{\omega^2}\sum_{\ell=0}^{\infty}\paren{2\ell+1}\paren{1-\left|\frac{A_{\omega\ell}^{\text{out}}}{A_{\omega\ell}^{\text{in}}}\right|^2}\ ,
\en
where the coefficients $A^{\text{out}}_{\omega\ell}$ and $A^\text{in}_{\omega\ell}$ are found through the partial waves analysis. Moreover, some important features can be recovered both at the high and low frequency limits. For the low frequency limit, the absorption cross section is related to the BH horizon area, $A_h=4\pi r_h^2$ \cite{Das1997}. Figure \ref{fig:phors} displays the horizon area as a function of $q$ for ModMax BHs in several configurations of $\beta$. It can be seen that the charge screening effect modifies the horizon area, i.e., for a given $q$, the horizon area grows as $\beta$ decreases.

Another interesting limiting behavior is that of the high-frequency domain. In this limit, the massless scalar waves can be approximated to the null geodesics studied before and the absorption cross section is just the capture cross section for a massless particle, which we shall denote $\sigma_\text{geo}$, and is given by 
\be \label{eq:sigmageo}
\sigma_{\text{geo}}=\pi b_c^2=\pi\frac{r_c^2}{f_c} \ .
\en
Moreover, the high-frequency limit also corresponds to the eikonal regime. In this regime, D\'{e}canini \emph{et al.} \cite{Decanini2011} 
shown that the oscillatory part of the absorption cross section can be adequately approximated as
\be \label{eq:sigmahf}
\sigma_{\text{hf}}\paren{\omega}\approx\sigma_{\text{geo}}+\sigma_\text{osc}\paren{\omega}\ ,
\en
where $\sigma_{\text{osc}}\paren{\omega}$ is the oscillating part of the absorption cross section for high frequencies and is given by the \emph{sinc approximation} which is given by the function 
\be \label{eq:sinc}
\sigma_{\text{osc}}\paren{\omega}=-8\pi^2\frac{\lambda}{\Omega_c^3}e^{-\pi\lambda/\Omega_c}\text{sinc}\paren{2\pi\omega/\Omega_c}\ ,
\en
where $\text{sinc}\paren{x}\equiv \sin(x)/x$ is the cardinal sine, $\Omega_c=\sqrt{f(r_c)}/r_c$ is the angular velocity of the circular null geodesic and $\lambda$ is the Lyapunov exponent associated with the null unstable geodesic on the photon sphere of the spacetime under scrutiny. It can be shown that this Lyapunov exponent is given by \cite{Cardoso2009} 
\be
\lambda=\sqrt{\frac{f_c}{2b_c}\paren{2-b_c^2f''_c}} \ .
\en

\begin{figure*}[t]
\centering
\includegraphics[width=0.9\textwidth]{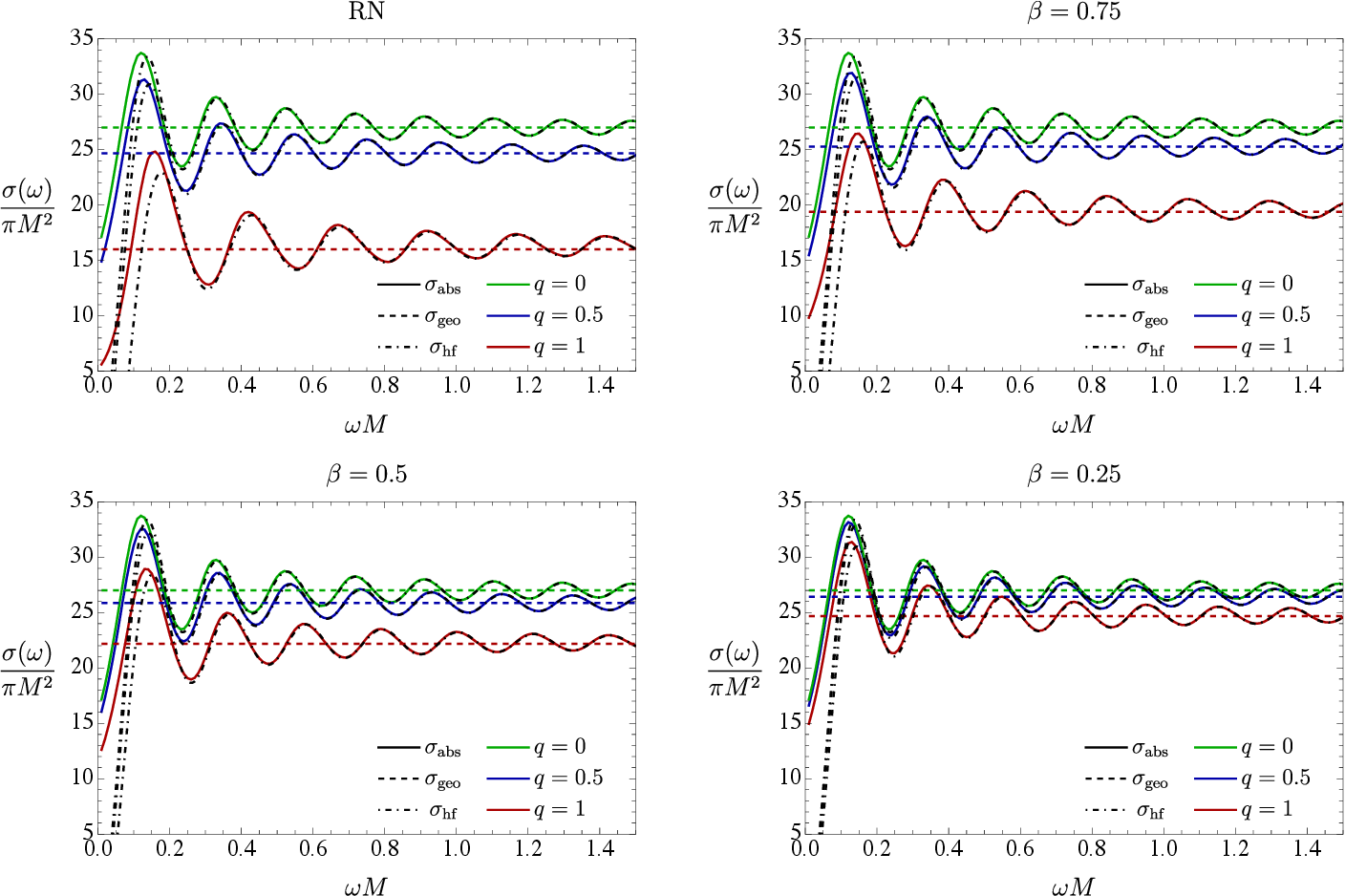}
\caption{Plot of the different absorption cross sections for different values of $\beta=e^{-\gamma}$ and different charges. Solid lines represent the total absorption cross section, $\sigma_{\text{abs}}$, obtained numerically. The dashed lines represent the geometric absorption cross section, $\sigma_\text{geo}$ and the dot-dashed lines are the high-frequency absorption cross section, $\sigma_{\text{hf}}$, given by the sinc approximation.}
\label{fig:pabsCS}
\end{figure*}

Figure \ref{fig:pabsCS} displays the absorption cross section for Einstein-ModMax BHs in different configurations of the parameter $\beta$. We display the total absorption cross section, calculated through the partial waves method, as well as the geometric cross section given by Eq. \eqref{eq:sigmageo} and the high-frequency absorption cross section corrected with the sinc approximation in Eq. \eqref{eq:sinc}. Notice that all these approximations are in good agreement with the numerical results within its domain of validity. Furthermore, the effect of nonlinearities of ModMax electrodynamics into the absorption cross section is reflected in the fact that absorption cross sections for different charges come closer as this nonlinearity parameter increases. This behavior is accentuated as $\beta$ becomes smaller and is due to the charge screening effect of this parameter. 
\section{Conclusions and perspectives} \label{sec:concl}

ModMax electrodynamics is an interesting theory of NLED due to the preservation of duality and conformal invariance. These features make the derivation of solutions in this theory much less cumbersome in contrast to other theories of NLED. Notwithstanding the apparent simplicity of this theory, it still offers interesting departures from Maxwell electrodynamics, being the vacuum birefringence among the most interesting ones, and interesting departures are expected when studying photon propagation in spacetimes dominated by ModMax electrodynamics \cite{GuzmanHerrera2024}.

In this work we have studied the scattering and absorption of massless scalar waves by a ModMax BH. As such, by applying the well known partial waves method, we have obtained the absorption and scattering cross section for these waves. Our results agree with those of Maxwell electrodynamics in the adequate limit (c.f. \cite{Crispino2009}), which validate the method that has been used. Moreover, we have also computed the analytical approximations for the scattering cross section via the classical scattering cross section obtained through geodesic analysis and the glory approximation. Also, for the absorption cross section we have computed the low and high-frequency approximations given by the horizon area and the sinc approximation, respectively. Both the scattering and absorption cross sections are in good agreement with these approximations in the adequate regimes, regardless of the value of $\beta$ (see Figs. \ref{fig:SCSgridcompact} and \ref{fig:pabsCS}).

The results for the scattering and absorption cross sections for ModMax electrodynamics are qualitatively similar to those for Maxwell electrodynamics. In fact, ModMax electrodynamics (in the static regime) has an overall effect of screening the charges, hence a growing nonlinearity parameter will be manifest by suppressing the effect of the charges. As such, it is expected that as nonlinearity grows (i.e., as $\beta$ decreases), the charge screening will become more important and the resulting cross sections for different charges will approximate to each other. This has been corroborated by our results.

Since we have considered the simple case of a massless scalar field minimally coupled to gravity, and not to the electromagnetic field, the results were somehow expected. However, the scenario may change significantly when studying the propagation of photons and electromagnetic waves in a ModMax BH background. Since this theory of NLED possesses vacuum birefringence, it is expected that departures from Maxwell's results will arise when considering the scattering and absorption of electromagnetic waves, we hope to address this problem in the future.

\begin{figure*}[t]
\centering
\includegraphics[width=0.9\textwidth]{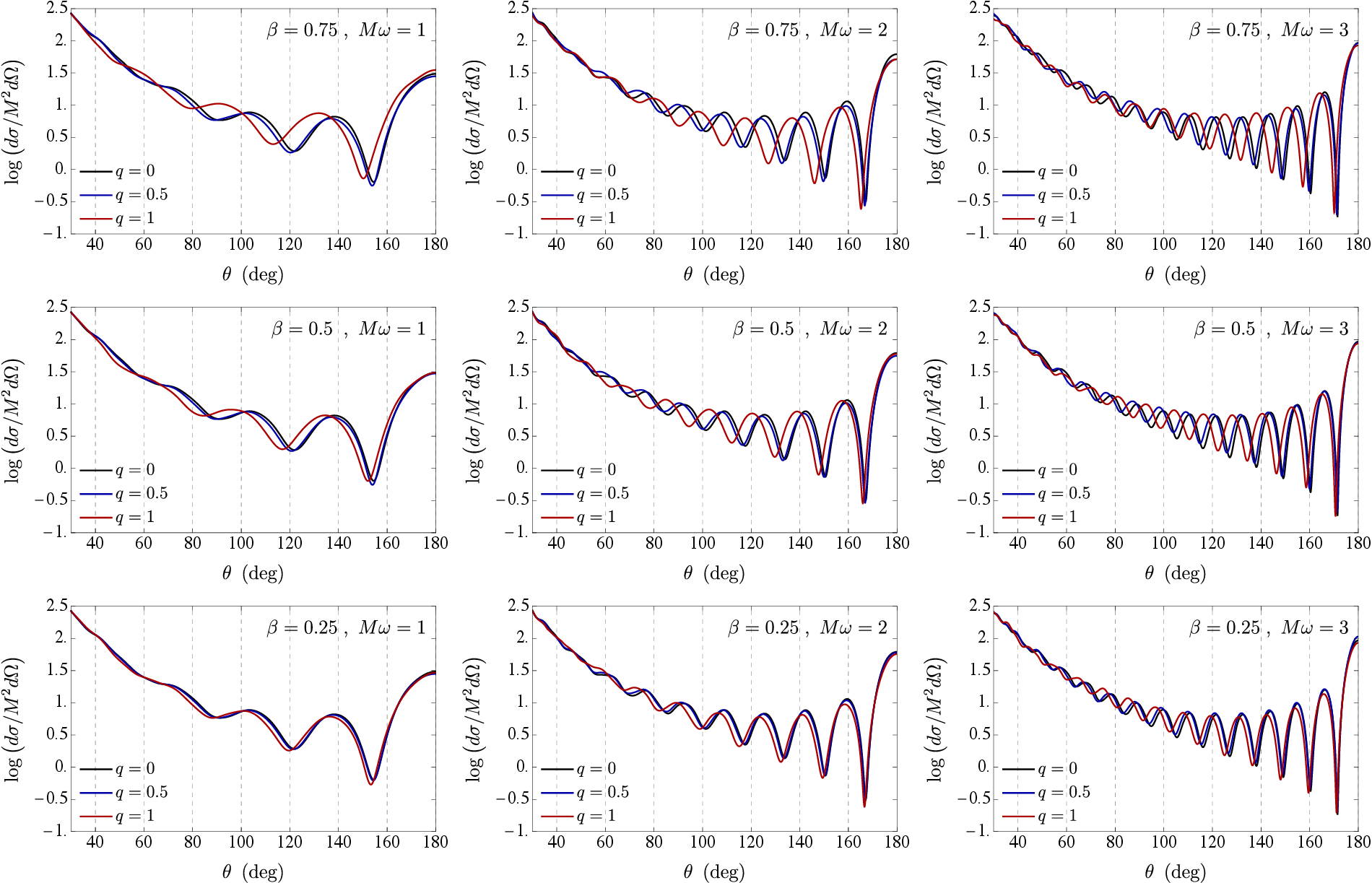}
\caption{Differential scattering cross section for Einstein-ModMax BHs for different values of $M\omega$, $q=Q/M$ and $\beta$. Each column displays a different value of $M\omega$, while the rows correspond to variations of $\beta$. 
The effect of the screening can be observed as the curves converge with decreasing $\beta$.}
\label{fig:apSCSdifbetas}
\end{figure*}

\begin{acknowledgments}
	This work was supported by the National Council for Scientific and Technological Development - CNPq and FAPERJ - Fundaçāo Carlos Chagas Filho de Amparo à Pesquisa do Estado do Rio de Janeiro, Processo SEI 260003/014960/2023 (MLP). MLP thanks CINVESTAV-IPN (Mexico City) for their hospitality, where part of this work was written.
\end{acknowledgments}




\bibliography{bibScatModMax}

\end{document}